\documentstyle[pre,aps,preprint,epsfig]{revtex}
\def\gtwid{\mathrel{\raise.3ex\hbox{$>$\kern-.75em\lower1ex\hbox{$\sim$}}}}
\def\ltwid{\mathrel{\raise.3ex\hbox{$<$\kern-.75em\lower1ex\hbox{$\sim$}}}}

\begin{document}

\draft
\tighten

\preprint{\vbox{\hfill quant-ph/0112139 \\
          \vbox{\hfill NSF-ITP-01-184} \\
          \vbox{\hfill December 2001} \\
          \vbox{\vskip0.5in}
         }}

\title{Sub-Planck Structure, Decoherence, and Many-Body Environments}

\author{
Andrew Jordan\footnote{E--mail: \tt ajordan@physics.ucsb.edu}
   and 
Mark Srednicki\footnote{E--mail: \tt mark@physics.ucsb.edu}
       }
\address{Department of Physics, University of California,
         Santa Barbara, CA 93106 
         \\ \vskip0.5in}

\maketitle

\begin{abstract}
\normalsize{
In a recent paper [Nature 412, 712 (2001)], Zurek has argued that
(1) time evolution typically causes chaotic quantum systems to generate
structure that varies on the scale of phase-space volume elements
of size $(\hbar^2/A)^d$, where $A$ is a classical action
characteristic of the state and $d$ is the number of degrees of
freedom, and that (2) this structure implies that a small change
in a phase-space coordinate $X$ by an amount 
$\delta X \sim \hbar X/A$ generically results in an orthogonal
state.  While we agree with (1), we argue that (2) is not correct
if the number of degrees of freedom is small.  
Our arguments are based on the Berry-Voros ansatz for the structure
of energy eigenstates in chaotic systems.  We find,
however, that (2) becomes valid if the number of degrees of
freedom is large.  This implies that 
many-body environments may be crucial for the phenomenon
of quantum decoherence.
}
\end{abstract}

\pacs{}

Zurek \cite{z} has recently pointed out that, contrary to prior
expectations \cite{bb}, a quantum system with $d$ degrees of freedom
will generically develop structure on the scale of phase-space
volume elements of size $(\hbar^2/A)^d$, where $A$ is a typical 
classical action associated with the state.  This is diagnosed
by considering the Wigner function for the quantum state $|\psi\rangle$,
\begin{equation}
W(x,p) = \int {d^ds\over(2\pi\hbar)^d}\,
\exp(ip\!\cdot\!s/\hbar)
\psi^*(x+{\textstyle{1\over2}}s)
\psi(x-{\textstyle{1\over2}}s),
\label{w}
\end{equation}
where $\psi(x)=\langle x|\psi\rangle$ is the position-space wave
function.  (We have assumed a pure state; the generalization to
mixed states is given in [1].)
By considering some simple superpositions of gaussian wave packets
(coherent states), Zurek shows that there can be fluctuations of $W(x,p)$
in the position $x$ on scales $\lambda_{x,\rm{min}}\sim\hbar/P$ (where $P$
is a typical classical value of the momentum $p$), and in $p$ on scales
$\lambda_{p,\rm{min}}\sim\hbar/L$ (where $L$
is a typical classical value of $x$).  The corresponding phase-space
volume is 
$(\lambda_{x,\rm{min}} \lambda_{p\rm{min}})^d\sim(\hbar^2/A)^d$,
where $A \sim PL$.

Zurek goes on to consider the possibility that this sub-Planck structure 
may have important physical significance if the system in question
serves as an environment for a qubit (or other two- or few-state
system).  In particular, he argues that if the two possible
states of the qubit cause slightly different
displacements in the phase space of the environment,
then these two possible environmental states will be nearly orthogonal.
This is then tantamount to decoherence of the quantum state of the
qubit by the environment.  To be specific, consider 
the action of a unitary displacement operator 
\begin{equation}
D(\delta p,\delta x)
=\exp[i(\delta p\!\cdot\!\hat x + \delta x\!\cdot\!\hat p)/\hbar],
\label{d}
\end{equation}
where $\hat x$ and $\hat p$ are the position and momentum operators,
on a state $|\psi\rangle$ of the environment.  The claim of [1]
is that
\begin{equation}
\langle\psi|D(\delta p,\delta x)|\psi\rangle \simeq 0 
\qquad\hbox{for}\qquad
\delta x \gtwid \hbar/P~~\hbox{or}~~\delta p \gtwid \hbar/L.
\label{d2}
\end{equation}
We will argue, however, that this is not correct in general
for systems with few degrees of freedom.
On the other hand, we find that eq.~(\ref{d2}) {\it is\/} generically
correct if the system (which is playing the role of the environment)
has many degrees of freedom.

Much of the analysis of [1] is based on the formula
\begin{equation}
\bigl|\langle\psi|D(\delta p,\delta x)|\psi\rangle\bigr|^2
= (2\pi\hbar)^d\int d^dx\,d^dp\;
W(x,p) W(x{+}\delta x,p{+}\delta p).
\label{dsq}
\end{equation}
However, a somewhat simpler and equally valid formula is
\begin{equation}
\langle\psi|D(\delta p,\delta x)|\psi\rangle 
= \int d^dx\,d^dp\;
e^{i(\delta p\cdot x + \delta x\cdot p)/\hbar}\,W(x,p).
\label{d3}
\end{equation}
Here we have made use of one of the basic properties of the Wigner
function: expectation values of symmetrically ordered operators
can be written as simple integrals.  Note that all $x$'s and $p$'s
in eq.~(\ref{d3}) are now classical integration variables.
Also, in the special case $\delta x=0$, eq.~(\ref{d3}) reduces
to eq.~(16) of [1].

From eq.~(\ref{d3}), we see immediately that we do not need to
have any information about the small-scale structure
in $W(x,p)$ in order to draw conclusions about the behavior of
$\langle D(\delta p,\delta x)\rangle$ for small 
$\delta p$ and $\delta x$.  This follows simply from the
mathematics of the Fourier transform, which relates small-scale structure
in the original variables ($x$ and $p$) to {\it large}-scale structure 
in the transformed variables ($\delta p$ and $\delta x$).
In fact, according to eq.~(\ref{d3}), the implication of the 
small-scale structure in $W(x,p)$ is that 
$\langle D(\delta x,\delta p)\rangle$ will be generically 
nearly zero only for
$\delta x \gtwid \hbar/\lambda_{p,\rm{min}} \sim L$ or
$\delta p \gtwid \hbar/\lambda_{x,\rm{min}} \sim P$.
That is, to get a generically orthogonal state,
the shift in coordinates or momenta must be comparable
to the {\it classical\/} scales that characterize the original state.

On the other hand,
what happens to $\langle D(\delta p,\delta x)\rangle$ at {\it small\/}
$\delta p$ or $\delta x$ is governed in turn by the behavior of
$W(x,p)$ for {\it large\/} $x$ or $p$.  If we suppose that the 
corresponding state has spread out over the available phase space
(which is characteristic of chaotic systems, and is explicitly assumed
in [1]), then we can treat the large-scale structure of 
the Wigner function via the Berry-Voros ansatz \cite{b,v1,v2}
\begin{equation}
W(x,p) \sim \delta(H(x,p)-E),
\label{bv}
\end{equation}
where $\delta(E)$ is the Dirac delta function, $H(x,p)$ is the
hamiltonian, and $E$ is the energy (assumed to be classically
sharp).  This ansatz simply corresponds to the ergodic distribution
that covers the constant-energy surface uniformly, according to
the Liouville probability measure.  It is devoid of the small-scale
structure elucidated in [1], but correctly describes the large-scale
structure.  When eq.~(\ref{bv})
is used in eq.~(\ref{d3}), the restriction of the range of
integration to a thin shell of phase space will produce ringing in 
$\langle D(\delta p,\delta x)\rangle$.  The first zeroes will
occur at $\delta x \sim \hbar/P$ and $\delta p \sim \hbar/L$;
these are the scales where, according to [1],
$\langle D(\delta p,\delta x)\rangle$ should become small.  
This is correct, but then the ringing effect implies that
there are recurrent maxima and minima in
$\langle D(\delta p,\delta x)\rangle$, with only a power-law 
fall-off of the peak heights.  (The exponents in these power laws depend
on the details of the system; more on this below.)  
This is contrary to the supposition of [1] that any recurrences
will be strongly suppressed in general (even though they occur
in the example of four superposed gaussians that is worked out explicitly).
Finally, for $\delta x \sim L$
and $\delta p \sim P$, the small-scale structure of $W(x,p)$
becomes important, and 
$\langle D(\delta p,\delta x)\rangle$ goes rapidly to zero.

To illustrate this, we consider a paradigmatic chaotic system,
a two-dimensional billiard such as the stadium \cite{stad}.
In this case $H(x,p)={1\over2m}p^2$ when $x$ is inside,
and $H(x,p)=\infty$ when $x$ is outside.  
If (for pedagogical simplicity)
we take the billiard to be nearly circular with radius $L$
(e.g, a stadium with a short straight segment \cite{li}, 
or a circle with rough walls \cite{rough}), and the bouncing ball
to have energy $E={1\over2m}P^2$, 
then from eqs.~(\ref{d3}) and (\ref{bv}) we get
\begin{equation}
\langle D(\delta p,\delta x)\rangle 
= J_0(P|\delta x|/\hbar)\,{J_1(L|\delta p|/\hbar)\over L|\delta p|/2\hbar},
\label{djj}
\end{equation}
where $J_n(\xi)$ is an ordinary Bessel function.
Here the oscillations and slow decay of the Bessel functions correspond
to the ringing phenomenon that we described above.  
Also, because the Berry-Voros ansatz does not
have the correct small-scale structure, we do not see the cutoffs
at $\delta p \sim P$ and $\delta x \sim L$ that would otherwise be there.

We can also go a little further than this for a chaotic system.  
Some time ago,
Berry \cite{b} conjectured that, in such a system, the energy
eigenfunctions (in position space)
have the statistical properties of a gaussian-random
superposition of plane waves, all with the wavenumber $k$ given locally
by classical energy conservation and the fundamental quantum relation
$k=P/\hbar$, but with wave-vectors pointing in random directions.  
Since then, this conjecture
has been supported by a number of numerical experiments \cite{mk,as,lr,ss},
and it underlies the successful characterization of the distribution
of conductance peak heights in quantum dots in the Coulomb-blockade
regime \cite{jsa,al,ms0}.  If we use an overbar to denote the statistical
random-wave average, then Berry's conjecture can be neatly summarized via
\begin{equation}
\overline{W(x,p)} \sim \delta(H(x,p)-E),
\label{wbar}
\end{equation}
plus the statistical statement that the wave-function $\psi(x)$ is
a gaussian random variable.  This in turn can be shown to imply \cite{ss}
\begin{equation}
\overline{W(x,p)^2} = {\overline{W(x,p)}}^{\,2} + O(\hbar^{d-1})
\label{w2bar}
\end{equation}
for $d\ge2$.  In this formula, 
$\overline{W(x,p)}$ itself is to be regarded as $O(\hbar^0)$.
Eq.~(\ref{w2bar}) means that each particular energy eigenfunction 
will have only
small deviations from eq.~(\ref{wbar}) on large scales.  Furthermore,
eqs.~(\ref{wbar}) and (\ref{w2bar}) also apply for generic time-dependent
superpositions of energy eigenstates, 
\begin{equation}
|\psi(t)\rangle 
= \sum_\alpha c_\alpha\,e^{-iE_\alpha t/\hbar}|\alpha\rangle,
\label{psi}
\end{equation}
and for generic mixed states,
\begin{equation}
\rho(t) = \sum_{\alpha\beta} c_{\alpha\beta}\,
                             e^{-i(E_\alpha-E_\beta)t/\hbar}
                             |\alpha\rangle\langle\beta|,
\label{rho}
\end{equation}
provided we assume that different eigenfunctions are statistically 
independent superpositions of random waves.  (Also, we assume that the
quantum energy uncertainty is classically small.) Thus,
in order to construct a state that is, for example, localized in 
a particular small region of phase space,
we must carefully choose the magnitudes {\it and phases\/} of the 
$c_\alpha$'s in eq.~(\ref{psi}) or the $c_{\alpha\beta}$'s in 
eq.~(\ref{rho}).  Once we have done this,
we expect that subsequent time evolution will eventually spread the packet
over the corresponding energy surface, its distribution on
large scales governed by the classical ergodic distribution.  
In quantum mechanics,
this time evolution is given entirely by the phase changes in
eq.~(\ref{psi}), and so these must be capable of producing this
spreading.  This is guaranteed by eqs.~(\ref{wbar}) and (\ref{w2bar}).
In a many-body system, this same effect corresponds to the approach
to thermal equilibrium \cite{ms1,ms2,ms3}.

We have shown that the small-scale structure in phase space discovered
in [1] does not generically result in small values of the overlap 
between a displaced state and the original unless the displacements
are classically large.  However, this conclusion does not take into
account new effects that can arise when $d$, the number of degrees of
freedom, is large.  Essentially, what happens for $d\gg 1$ is that
the power-law exponent for the fall-off of the recurrent peaks in  
$\langle D(\delta p,\delta x)\rangle$ becomes large with $d$,
suppressing all the recurrences, and restoring the conclusions of [1].
We now turn to our exposition of this phenomenon.

For definiteness, we will consider a dilute
gas of $N$ hard spheres in a three-dimensional box of
volume $L^3$.  This system can also be viewed as
a single bouncing ball in a complicated $3N$-dimensional chaotic billiard,
and so we will assume that Berry's random-wave conjecture holds for
it at sufficiently high total energy $E$ \cite{ms1}.  We take 
$E={N\over 2m}P^2$, so that $P$ is the typical
momentum of a single gas particle.  The appropriate
generalization of eq.~(\ref{djj}) is then
\begin{equation}
\langle D(\delta p,\delta x)\rangle 
= { J_\nu(N^{1/2}P|\delta x|/\hbar) \over 
(N^{1/2}P|\delta x|/2\hbar)^\nu \nu!}
\,\prod_{i=1}^{3N}{\sin(L\,\delta p_i/2\hbar)\over 
L\,\delta p_i/2\hbar},
\label{djj2}
\end{equation}
where $\nu=(3N-2)/2$.
To perform the position-space integral that results in the multiple product, 
we ignored
the spatial regions that are excluded by the requirement that the
hard-sphere centers not come closer than the hard-sphere diameter;
this should not be an important effect for a dilute gas.
In the large-$N$ limit, the $\delta x$ dependent factor in 
eq.~(\ref{djj2}) becomes $\exp(-P^2|\delta x|^2/6\hbar^2)$.  
If we assume that the 
$\delta p_i$'s are all of comparable magnitude, 
and that each is less than $\hbar/L$,
then the $\delta p$ dependent factor becomes 
$\exp(-L^2|\delta p|^2/24\hbar^2)$.
Thus we have
\begin{equation}
\langle D(\delta p,\delta x)\rangle 
= \exp(-P^2|\delta x|^2/6\hbar^2) 
  \exp(-L^2|\delta p|^2/24\hbar^2).
\label{djj3}
\end{equation}
From this formula we see that
$\langle D(\delta p,\delta x)\rangle$ goes rapidly to zero for 
either $|\delta x| \gtwid \hbar/P$ or
$|\delta p| \gtwid \hbar/L$, which is now in accord with [1].

Before concluding, we briefly return to systems with few degrees of
freedom, and consider the reconciliation of eqs.~(\ref{dsq}) and
(\ref{d3}).  Assuming that the state spreads out over the available
phase space, we have seen that eq.~(\ref{d3}) implies that
$\langle D(\delta p,\delta x)\rangle$ rings, with recurrent
zeroes at multiples of 
$\delta x \sim \hbar/P$ and $\delta p \sim \hbar/L$. 
How can this behavior be reproduced by eq.~(\ref{dsq})?
It is clear that this is possible if and only if $W(x,p)$
in fact has the small-scale structure identified in [1].
Thus there is an intimate relation between the large- and
small-scale structure of the Wigner function.

To summarize, we find that
small displacements $\delta x \gtwid \hbar/P$ in position
and/or $\delta p \gtwid \hbar/L$ of the quantum state of a
many-body chaotic system generically produce a nearly
orthogonal state, provided the system has many degrees
of freedom.  For a chaotic system with few degrees of
freedom, we find that the overlap between the displaced
and the original state has a series of zeroes spaced
by $\delta x \sim \hbar/P$ and $\delta P \sim \hbar/L$,
separated by broad peaks whose heights fall off as a power
of $\delta x$ or $\delta p$, with a final exponential
cutoff at $\delta x \sim L$ and $\delta p \sim P$.
These results imply that a many-body system, thought of
as an environment, will (not
unexpectedly) be more effective in causing a simple
system (such as a qubit) to undergo decoherence.

\begin{acknowledgments}

We thank Wojciech Zurek for discussions of his results.
M.S.~thanks the Institute for Theoretical Physics for
hospitality.
This work was supported in part by the National Science
Foundation under Grant Nos.~PHY00--98395 and PHY99-07949.

\end{acknowledgments}

\end{document}